\documentclass[fleqn,usenatbib]{mnras}
\usepackage{newtxtext,newtxmath}
\usepackage[T1]{fontenc}
\usepackage{ae,aecompl}

\usepackage{graphicx}
\usepackage{amssymb}
\usepackage{wasysym}
\usepackage{gensymb}
\usepackage[para,online,flushleft]{threeparttable}

\newcommand{\psr}{J2322$-$2650}
\newcommand{\psrb}{B1257+12}
\newcommand{\psrd}{J1719$-$1438}
\newcommand{\psre}{J0636+5128}

\title[PSR J2322--2650]{PSR J2322--2650 -- A low-luminosity millisecond pulsar \\with a planetary-mass companion}  
\author[R.~Spiewak et al.]{
\!\!R.~Spiewak,$^1$\thanks{E-mail: rspiewak@swin.edu.au (RS)} %alphabetical after here
M.~Bailes,$^{1,2}$ 
E.~D.~Barr,$^3$ 
N.~D.~R.~Bhat,$^4$ 
M.~Burgay,$^5$ 
\newauthor A.~D.~Cameron,$^3$ 
D.~J.~Champion,$^3$ 
C.~M.~L.~Flynn,$^1$ 
A.~Jameson,$^{1,6}$
S.~Johnston,$^7$ 
\newauthor M.~J.~Keith,$^8$ 
M.~Kramer,$^{3,8}$ 
S.~R.~Kulkarni,$^{9}$ 
L.~Levin,$^8$ 
A.~G.~Lyne,$^8$
V.~Morello,$^8 $
\newauthor C.~Ng,$^{10}$ 
A.~Possenti,$^5$ 
V.~Ravi,$^{9}$ 
B.~W.~Stappers,$^8$ 
W.~van Straten,$^{11}$ 
\newauthor C.~Tiburzi$^{3,12}$ \\  \\
$^1$Centre for Astrophysics and Supercomputing, Swinburne University of Technology, PO Box 218, Hawthorn, VIC 3122, Australia \\
$^2$ARC Centre of Excellence for Gravitational Wave Discovery (OzGrav), Mail H29, Swinburne University of Technology, PO Box \\ 218, Hawthorn, VIC 3122, Australia \\
$^3$Max-Planck-Institut f\"ur Radioastronomie, Auf dem H\"ugel 69, D-53121 Bonn, Germany \\
$^4$International Centre for Radio Astronomy Research, Curtin University, Bentley, WA 6102, Australia \\
$^5$INAF-Osservatorio Astronomico di Cagliari, via della Scienza 5, I-09047 Selargius, Italy \\
$^6$ARC Centre of Excellence for All-Sky Astronomy (CAASTRO), Mail H30, Swinburne University of Technology, PO Box 218,\\ Hawthorn, VIC 3122, Australia \\
$^7$CSIRO Astronomy and Space Science, Australia Telescope National Facility, PO Box 76, Epping, NSW 1710, Australia \\
$^{8}$Jodrell Bank Centre for Astrophysics, University of Manchester, Alan Turing Building, Oxford Road, Manchester M13 9PL, UK \\
$^{9}$Cahill Center for Astronomy and Astrophysics, MC 249-17, California Institute of Technology, Pasadena, CA 91125, USA  \\ 
$^{10}$Department of Physics and Astronomy, University of British Columbia, 6224 Agriculture Road, Vancouver, BC V6T 1Z1, Canada \\
$^{11}$Institute for Radio Astronomy \& Space Research, Auckland University of Technology, Private Bag 92006, Auckland 1142, New\\ Zealand \\
$^{12}$Fakult\"at f\"ur Physik, Universit\"at Bielefeld, Postfach 100131, D-33501 Bielefeld, Germany
}

\date{Accepted XXX. Received YYY; in original form ZZZ}
\pubyear{2017}

\begin{document}
\label{firstpage}
\pagerange{\pageref{firstpage}--\pageref{lastpage}}
\maketitle

\begin{abstract}
We present the discovery of a binary millisecond pulsar (MSP), PSR~J2322$-$2650, found in 
the Southern section of the High Time Resolution Universe survey.  
This system contains a 3.5-ms pulsar with a $\sim10^{-3}$\,M$_{\odot}$ companion in a 7.75-hour circular orbit.   
Follow-up observations at the Parkes and Lovell telescopes have led to precise measurements of the astrometric and spin parameters, including the period derivative, timing parallax, and proper motion. 
PSR\,\psr\ has a parallax of $4.4\pm1.2$\,mas, and is thus at an inferred distance of $230^{+90}_{-50}$\,pc, making this system a candidate for optical studies. 
We have detected a source of $R\approx26.4$\,mag at the radio position in a single $R$-band observation with the Keck Telescope, and this is consistent with the blackbody temperature we would expect from the companion if it fills its Roche lobe. 
The intrinsic period derivative of PSR~\psr\ is among the lowest known, $4.4(4)\times10^{-22}$\,s\,s$^{-1}$, implying a low surface magnetic field strength, $4.0(4)\times10^7$\,G. 
Its mean radio flux density of 160\,$\mu$Jy combined with the distance implies that its radio luminosity is the lowest ever measured, $0.008(5)$\,mJy\,kpc$^2$. 
The inferred population of these systems in the Galaxy may be very significant, suggesting that this is a common MSP evolutionary path. 
\end{abstract}


\section{Introduction}\label{sec:int}
Since the discovery of pulsars \citep{hbp+68}, more than 2500 have been detected with a wide range of spin periods and magnetic field strengths.  The majority of known pulsars are isolated, but roughly 10\,percent have companions with masses ranging from ($\sim10^{-6}$ - $\sim10^{1}$)\,M$_{\odot}$.

At irregular intervals, new types of pulsars %
are discovered that lead to breakthroughs in our understanding of theories of relativistic gravity or the pulsar emission mechanism, or how pulsars evolve. 
For example, the discovery of the double pulsar led to new tests of General Relativity \citep{bdap+03,lbk+04}, and the discovery of intermittent pulsars demonstrated that a radio pulsar's emission mechanism could exhibit bimodal behaviour \citep{klob+06}. 

After the discovery of the first binary pulsar, PSR~B1913+16 \citep{ht75}, also known as the Hulse-Taylor pulsar, \citet{bkk76} described a possible course of evolution of the system through an X-ray bright phase, during which the magnetic field of the pulsar is weakened and the pulsar's spin period reduced. 
When the first millisecond pulsar (MSP) was discovered by \citet{bkh+82}, \citet{acrs82} proposed an evolutionary track for ordinary pulsars to be spun up to millisecond periods by mass transferred from a binary companion, listing low mass X-ray binaries (LMXBs) among the possible progenitors. In the intervening 35 years, this has become the standard model for MSP production \citep[see, e.g.,][]{deloye08}, and some systems have been observed to transition between the LMXB and radio MSP states (e.g., PSR\,J1227$-$4853; \citealt{rrb+15}), providing support for the model proposed by \citet{acrs82}.
In this model, the mass of a neutron star's companion largely determines the final spin period of the recycled pulsar. 
Low-mass companions lead to MSPs (isolated or with white dwarf (WD) companions), whereas higher mass stars may themselves create a neutron star, leading to a system resembling the Hulse-Taylor 
pulsar.

The discovery of planets orbiting a pulsar (PSR\,B1257+12; \citealt{wf92}) challenged theorists to explain the formation of such systems, as did the discovery of the ``diamond planet" pulsar (PSR\,J1719$-$1438; \citealt{bbb+11a}). 
In fact, of the 2613 pulsars in the ATNF pulsar catalogue (v.1.56; \citealt{mhth05}), only 4 in the field have planetary-mass companions (defined as having masses less than $10^{-2}$\,M$_{\odot}$): PSRs~J0636+5128\footnote{\label{note:psre}Originally, PSR\,\psre\ was published by \citet{slr+14} as PSR\,J0636+5129, but the designation has been corrected by Arzoumanian et al., in prep.} \citep{slr+14}, 
B1257+12, J1311$-$3430 \citep{pgf+12a}, and J1719$-$1438. These pulsars are all MSPs, around which it is comparatively easy to detect low-mass companions via pulsar timing \citep[see][]{wolszczan97}, whereas no such low-mass companions have been detected around young pulsars \citep{kjhs15}. 
Various hypotheses have been proposed for the formation of the above systems, ranging from near-complete ablation of a companion, to the inheritance of planets formed around a main sequence star before the formation of the pulsar, to the development of planets in supernova fallback disks around young pulsars. These models and their implications have been discussed in several papers \citep[e.g.,][]{mlp16,mh01,wkb07}. 
Discoveries of new pulsars with planetary-mass companions are needed to expand our knowledge of the evolutionary scenarios and to discriminate among them.

The High Time Resolution Universe (HTRU) pulsar survey is a highly successful pulsar survey, which uses the Multibeam receiver on the Parkes telescope \citep{sswb+96} to observe the Southern sky \citep{kjvs+10}, with the Northern sky covered by the Effelsberg 100-m Radio Telescope in Germany \citep{bck+13}. 
To date, 996 pulsars have been detected in the Southern part of HTRU, of which 171 are new discoveries, according to the ATNF pulsar catalogue.

In this work, we define an MSP as a pulsar with rotational period less than 20\,ms and spin-down rate less than $10^{-17}$\,s\,s$^{-1}$.  
When deriving companion masses, if the pulsar mass is not known, we adopt the standard value of 1.4\,M$_{\astrosun}$.  
The layout of the paper is as follows. 
In \S~\ref{sec:obs}, we give an overview of the discovery of PSR~\psr, and describe follow-up timing and optical observations.  In \S~\ref{sec:pars}, we describe the system parameters found through timing and compare this with properties of other known pulsars.  In \S~\ref{sec:disc}, we look at how the system compares with other pulsars with planetary-mass companions and postulate possible formation scenarios for this system, and, finally, we offer some general conclusions in \S~\ref{sec:con}.

\section{Timing Observations}\label{sec:obs}
\subsection{Discovery of PSR~J2322$-$2650} \label{ssec:disc}
PSR~\psr\ was discovered in the HTRU high-latitude survey 
with Parkes on 2011 May 4 in a 285-second observation at 1400\,MHz. 
The initial detection had a signal-to-noise ratio of $\approx12$, and the source was confirmed with observations (starting July 2012) with the Lovell Telescope at the Jodrell Bank Observatory (JBO) at a centre frequency of 1520\,MHz.  At the time of the initial detection, the flux density was $\approx0.27$\,mJy (from the radiometer equation and taking into account the offset from boresight).  
The pulsar had a period of 3.463\,ms and a dispersion measure (DM) of 6.18\,pc\,cm$^{-3}$ in the discovery observation.  Follow-up observations soon revealed an orbit with a period of 7.75\,h and projected semi-major axis of only 0.0028\,lt-s.  

\subsection{Timing programs}\label{ssec:time}

\begin{table*}
 \centering 
 \caption{Follow-up observations of \psr\ -- receiver information \label{tab:obs}}
 \begin{threeparttable}
 \begin{tabular*}{2\columnwidth}{l l l c c c l}
  \hline
  Telescope & Receiver & Backend & Centre Frequency & Recorded BW & Obs.~Used/Recorded & Dates  \\ 
   & & & (MHz) & (MHz) & & (MJD) \\ \hline
  Lovell & & ROACH & 1520 & 512 & 239/279 & 56129-57848\\ \hline 
  Parkes & MB & CASPSR & 1382 & 400 & 30/44 & 56174-57261, 57761-57870 \\
  & & DFB3\tnote{a} & 1369 & 256 & 2/23 & 56156-56739 \\
  & & DFB4\tnote{a} & & & 2/21 & 56953-57341, 57761-57823 \\
  & H-OH & CASPSR & 1400 & 400 & 8/9 & 57472-57703 \\
  & & DFB4\tnote{a} & 1369 & 256 & 0/7 & 57621-57703 \\
  & 10/50cm & CASPSR & 728 & 200 & 0/2 & 56511, 57846 \\
  & & DFB3 & 732 & 64 & 0/2 & 56504, 56511 \\
  & & DFB4 & 3100 & 1024 & 2/3 & 56504, 56511, 57846 \\ \hline
 \end{tabular*}
 \begin{tablenotes}
 \item[a] CASPSR observations preferentially used where overlapping with DFB data in the same band\\
 \end{tablenotes}
 \end{threeparttable}
\end{table*}

Follow-up timing of \psr\ was carried out using the Parkes and Lovell telescopes, as described in Table~\ref{tab:obs}.  The timing data from Parkes (project ID P789) span a period of 4.8\,years, from MJD 56174 to 57846, with multiple receivers and pulsar processing systems. 
The majority of the observations at Parkes use the Multibeam (MB) receiver, which has a frequency range of 1220-1520\,MHz and cold-sky system equivalent flux density of 29\,Jy (for the centre beam\footnote{\url{http://www.atnf.csiro.au/research/multibeam/lstavele/description.html}}).  We used the H-OH receiver when the MB system was not available (2016 March 25 to Nov.~11), and the 10/50cm receiver for greater spectral coverage. The backends used are the ATNF digital filterbanks (DFBs) and CASPSR\footnote{CASPER Parkes Swinburne Recorder; \url{http://www.astronomy.swin.edu.au/pulsar/?topic=caspsr}}. 
The CASPSR backend coherently dedisperses the data, whereas the DFB backends do not, although, for a pulsar with such a small DM, this makes little practical difference. 
Observations with the Lovell Telescope cover the MJD range 56129 to 57848 and make use of a cryogenically cooled dual-polarization receiver with optimal performance in the frequency range 1350-1700\,MHz. The cold-sky system equivalent flux density of the system is 25\,Jy. The ROACH-based backend\footnote{\url{https://casper.berkeley.edu/wiki/ROACH}} 
Nyquist-samples the 512-MHz-wide band at 8-bit resolution and divides the band into 32$\times$16\,MHz wide sub-bands \citep{bjk+16}. Each sub-band is coherently dedispersed and folded in real time with the resultant pulse profiles stored with 1024 bins across the pulse profile. The sub-bands are combined in offline processing and, with the removal of known radio frequency interference (RFI) signals, a total bandwidth (BW) of approximately 384\,MHz is used. 

\begin{figure}
 \centering
 \includegraphics[width=\columnwidth]{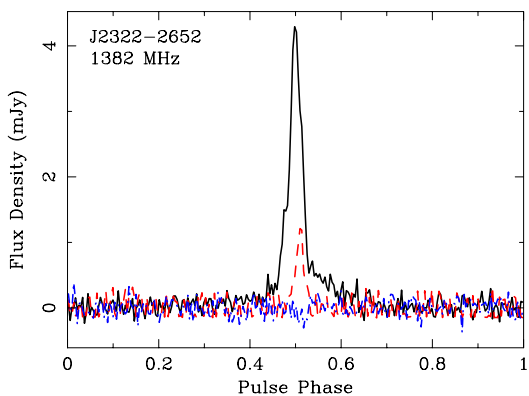}
 \caption{Integrated pulse profile from summed observations (equivalent integration time $\sim 50$\,ks) with linear (red dashed line) and circular (blue dash-dotted line) polarizations. The profile is well-approximated by a small number of Gaussian components and has a FWHM of 3\,percent.} 
 \label{fig:prof}
\end{figure}

Data from the Parkes observations were calibrated for flux and polarization information using separate observations of Hydra A from the Parkes P456 project. The data from the Lovell Telescope are not flux- or polarization-calibrated, but these effects are negligible for timing purposes, given the low polarization fraction. Figure~\ref{fig:prof} shows the integrated profile from the sum of several observations (to an equivalent integration time of $\sim50$\,ks) performed using the MB system and the CASPSR backend. 
At 1400\,MHz, the mean flux density is $S_{1400} = 0.16(2)$\,mJy (with the mean measurement uncertainty). 
The low signal-to-noise of the observations and significant instrumental effects make determination of the rotation measure difficult, even with the summed observations. 
Similarly, accurate measurement of the polarization position angle across the pulse is not possible. 
However, as the pulse profile is narrow (FWHM $=0.11(1)$\,ms at 1400\,MHz), precision timing is still possible.  
Due to interstellar scintillation, only $\sim 5$\,percent of the flux-calibrated timing observations have flux density $S_{1400}\gtrsim0.27$\,mJy, the flux density at the time of discovery.  \psr\ was not detected in 4 observations at 700\,MHz with the Parkes 10/50\,cm receiver, implying a flux density of $S_{700}\lesssim0.1$\,mJy for those epochs.  In observations at 3100\,MHz, the pulsar was detected at low significance, giving an estimated flux density of $\approx0.06$\,mJy (from one flux-calibrated observation).  
Given the limited number of observations at frequencies other than 1400\,MHz, accurate calculation of the spectral index was not possible.  
Observations at all three frequencies taken on the same day (with S/N of 3.2\footnote{The non-detection at 700\,MHz was scaled to a S/N of 5.0 to calculate the upper limit on the flux density for the spectral index.}, 13.6, and 9.0 at 700\,MHz, 1400\,MHz, and 3100\,MHz, respectively) imply a spectral index of $\approx -0.5$, although this is subject to bias due to scintillation.

\section{System Parameters}\label{sec:pars}

\begin{table}
 \caption{Pulsar parameters from radio timing using {\sc tempo2} -- uncertainties on direct timing parameters from {\sc tempo2}\label{tab:par}}
 \begin{threeparttable}
 \begin{tabular*}{\columnwidth}{l c}
  \hline 
  Parameter & Value \\ \hline 
  Right Ascension (J2000) (h:m:s) & 23:22:34.64004(3) \\
  Declination (J2000) (d:m:s) & $-$26:50:58.3171(6) \\
  Period, $P$ (s) & 0.00346309917908790(11) \\
  Period derivative, $\dot P$ (s\,s$^{-1}$) & $5.834(15)\times10^{-22}$ \\
  Period epoch\tnote{a} (MJD) & 56152.0 \\
  DM (pc\,cm$^{-3}$) & 6.149(2) \\
  Parallax (mas) & 4.4(12) \\
  Proper motion in RA (mas yr$^{-1}$)& $-2.4(2)$ \\
  Proper motion in Dec (mas yr$^{-1}$)& $-8.3(4)$\\
  Binary model & ELL1 \\
  $P_{\rm b}$ (d) & 0.322963997(6) \\
  $T_{\rm ASC}$ (MJD) & 56130.35411(2) \\
  $x$ (lt-s) & 0.0027849(6) \\
  $\epsilon_1$ & $-0.0002$(4) \\
  $\epsilon_2$ & 0.0008(4) \\
  $\dot P_{\rm b}$\tnote{b} (s\,s$^{-1}$) & $\lesssim6\times10^{-11}$ \\
  $\dot x$\tnote{b} (lt-s\,s$^{-1}$) & $\lesssim3\times10^{-14}$ \\
  Data span (yr) & 4.8 \\ 
  Weighted RMS residual ($\mu$s) & 7.3 \\
  Number of TOAs & 338 \\ 
  $S_{1400}$ (mJy) & 0.16(2) \\ 
  FWHM at 1.4\,GHz (ms) & 0.11(1) \\ \hline  
  \multicolumn{2}{c}{Derived Parameters} \\ \hline
  $B_{\rm surf}$ (G) & $4.548(12)\times10^7$ \\ 
  Parallax-derived distance (kpc) & $0.23^{+0.09}_{-0.05}$ \\ %or 0.23(6)
  DM-derived distance\tnote{c} (kpc) & 0.76 \\  
  $\mu_{\rm Tot}$ (mas yr$^{-1}$)& 8.6(4) \\
  $V_{\rm trans}$\tnote{d} (km\,s$^{-1}$) & 20(5) \\ %$19.4\pm1.8$
  $\dot P_{\rm int}$\tnote{e} (s\,s$^{-1}$) & $4.4(5)\times10^{-22}$ \\
  $\dot E_{\rm int}$ (erg\,s$^{-1}$) & $4.2(4)\times10^{32}$ \\ 
  $e$\tnote{b} & $\lesssim0.0017$ \\  % 
  $\omega$ (deg) & 333(27) \\
  Predicted $\dot \omega$\tnote{f} (deg\,yr$^{-1}$) & 1.6 \\ 
  Mass function (M$_{\odot}$) & 2.229(1)$\times10^{-10}$ \\
  Min.~companion mass\tnote{f} (M$_{\odot}$) & 0.0007588(2) \\
  Min.~companion density (g\,cm$^{-3}$) & 1.84\\%
  $L_{1400}$\tnote{e} (mJy\,kpc$^2$) & 0.008(5) \\ \hline
 \end{tabular*}
 \begin{tablenotes}
 \item[a] Period Epoch also used as Position Epoch and DM Epoch\\
 \item[b] 2-$\sigma$ upper limit\\
 \item[c] YMW16 model \citep{ymw17}\\
 \item[d] With respect to the Local Standard of Rest\\
 \item[e] Using parallax-derived distance \\
 \item[f] Assuming a pulsar mass of 1.4\,M$_{\odot}$\\
 \end{tablenotes}
 \end{threeparttable}
\end{table}
% data taken from J2322-2652_paper.par on July 4 -- updated to _paperR.par on Sept. 4
% using J2322-2652_paperR.tim

\begin{figure}
 \centering
 \includegraphics[width=\columnwidth]{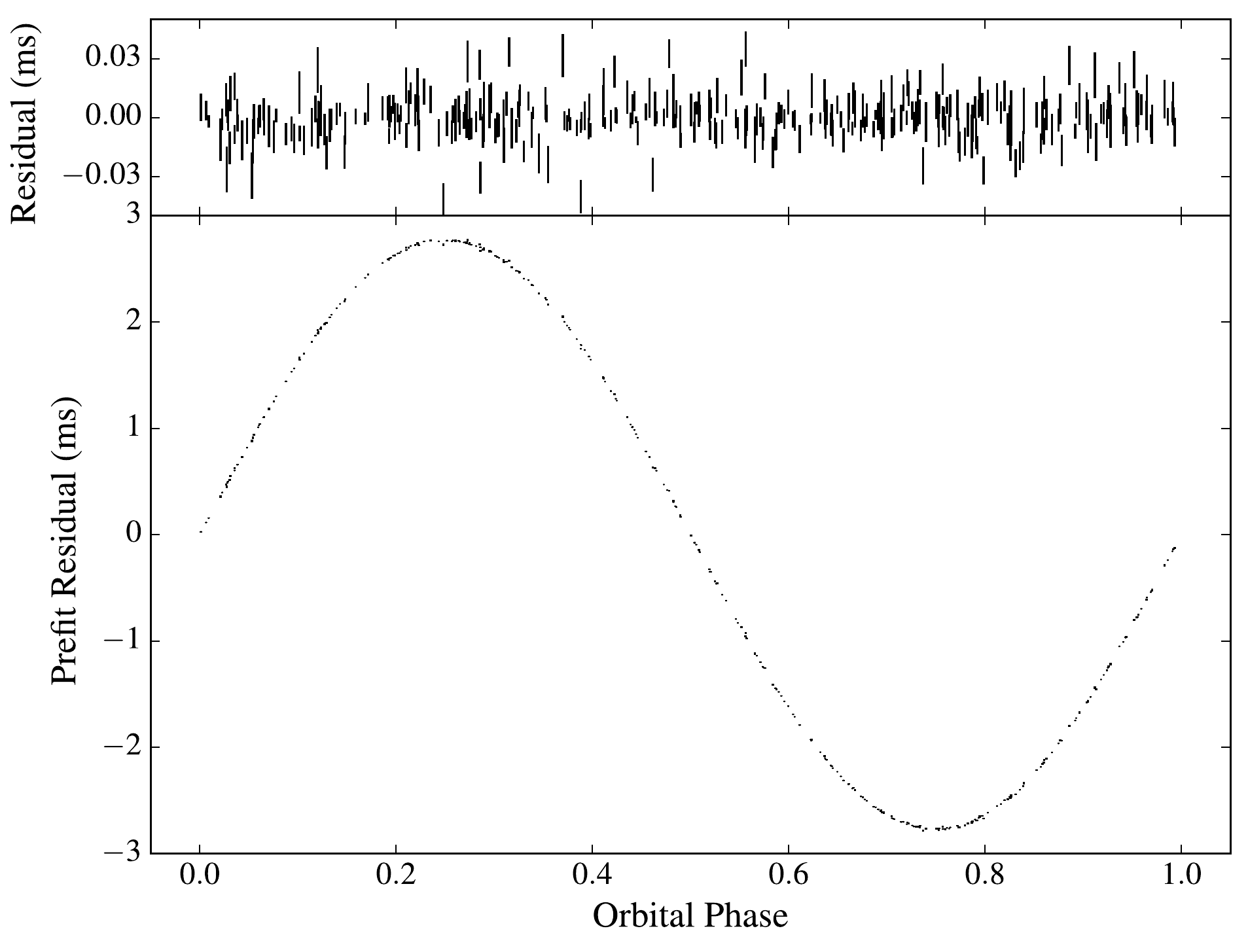}
 \caption{Upper panel: Pulse timing residuals for \psr\ with the optimal parameters (listed in Table~\ref{tab:par}). Lower panel: Residuals before fitting for the semi-major axis, demonstrating the effect of the binary motion. There is no significant orbital eccentricity, nor is there evidence for eclipses or excessive dispersive delays at superior conjunction of the pulsar (orbital phase 0.25). \label{fig:orb}}
\end{figure}

Using the \textsc{tempo2} software package \citep{hem06} with the `ELL1' binary model (Wex, N., unpublished.), the combined data from Parkes and JBO result in a weighted RMS timing residual of 7.2\,$\mu$s. The timing data and resulting parameters are available in the online journal. Table~\ref{tab:par} shows the parameters
of the timing solution covering the entire data span, with
the nominal $1\sigma$ uncertainties resulting from the fit. The derived
parameters are also reported. 
2-$\sigma$ upper limits are determined for the time derivatives of 
orbital period ($\dot P_{\rm b}$) and projected semi-major axis ($\dot x$) by fitting for the parameters individually to determine the uncertainties. 
The `ELL1' binary model uses the epoch of ascending node, $T_{\rm ASC}$, and the first and second Laplace-Lagrange parameters, $\epsilon_1 = e\sin(\omega)$ and $\epsilon_2 = e\cos(\omega)$, where $e$ is the eccentricity and $\omega$ is the orbital longitude. Figure~\ref{fig:orb} shows the effect of the binary orbit in the timing residuals.

\subsection{Astrometry} \label{ssec:dist} 
From the timing parallax of $4.4\pm1.2$\,mas, we infer a distance of only $230^{+90}_{-50}$\,pc. That is the reference value used throughout this paper as the correction for the Lutz-Kelker bias 
($220^{+100}_{-50}$\,pc according to the formula in \citealt{vwc+12}) is negligible given the current uncertainty on the parallax. 
On the other hand, the latest electron density model, YMW16 \citep{ymw17}, suggests a distance of 760\,pc. Using the NE2001 model \citep{cl02}, we find $d = 320$\,pc, which is consistent with the parallax distance, whereas the YMW16 distance is not. Typical uncertainties for DM-derived distances are $\sim20$ to 30\,percent. The magnitude of this discrepancy in the electron density models is not uncommon for nearby pulsars.

As of May 2017, the parallax measurement is significant at the $>3$-$\sigma$ level, but continued timing will allow for improved precision.  At a distance of 230\,pc, \psr\ is closer than all but 7 pulsars (1 MSP) with consistent distance measurements\footnote{``Consistent'' distances, from the ATNF pulsar catalogue, are those from timing parallax or independent distance measurements, or where the YMW16 and NE2001 models agree within a factor of 3: 2396 pulsars (139 MSPs) total.}. The 2-$\sigma$ upper limit on the parallax value gives a lower limit on the distance of 150\,pc. 

The total proper motion, from timing, is $\mu_{\rm Tot} = 8.6(4)$\,mas\,yr$^{-1}$, which, combined with the parallax distance, gives a transverse velocity of 10(3)\,km\,s$^{-1}$.  Converting this to the Local Standard of Rest yields a transverse velocity of 20(5)\,km\,s$^{-1}$.

\subsection{Intrinsic properties} \label{ssec:pdot} 
Our timing yields an observed period derivative of $\dot P_{\rm obs} = 5.834(15)\times10^{-22}$\,s\,s$^{-1}$, which implies a magnetic field strength of just $B_{\rm surf} = 4.55(1)\times10^7$\,G (where the given uncertainty does not take into account the assumptions made in the derivation). 
Correcting for the Shklovskii effect (ignoring the negligible contribution of the Galactic potential), we find an intrinsic period derivative of $\dot P_{\rm int} = 4.4(5)\times10^{-22}$\,s\,s$^{-1}$. 
This is the lowest, significant intrinsic $\dot P$ currently known after correcting for the Shklovskii effect, with the uncertainty derived from the large uncertainty on the parallax distance and the small uncertainty on the observed period derivative. 
If we assume $\dot P_{\rm int}$ must be positive, the distance is constrained to 
be $\lesssim0.9$\,kpc. The 2-$\sigma$ upper limit from the parallax corresponds to a distance of $>150$\,pc and $\dot P_{\rm int,max} = 4.9\times10^{-22}$\,s\,s$^{-1}$. 
 
\begin{figure}
 \centering
 \includegraphics[width=\columnwidth]{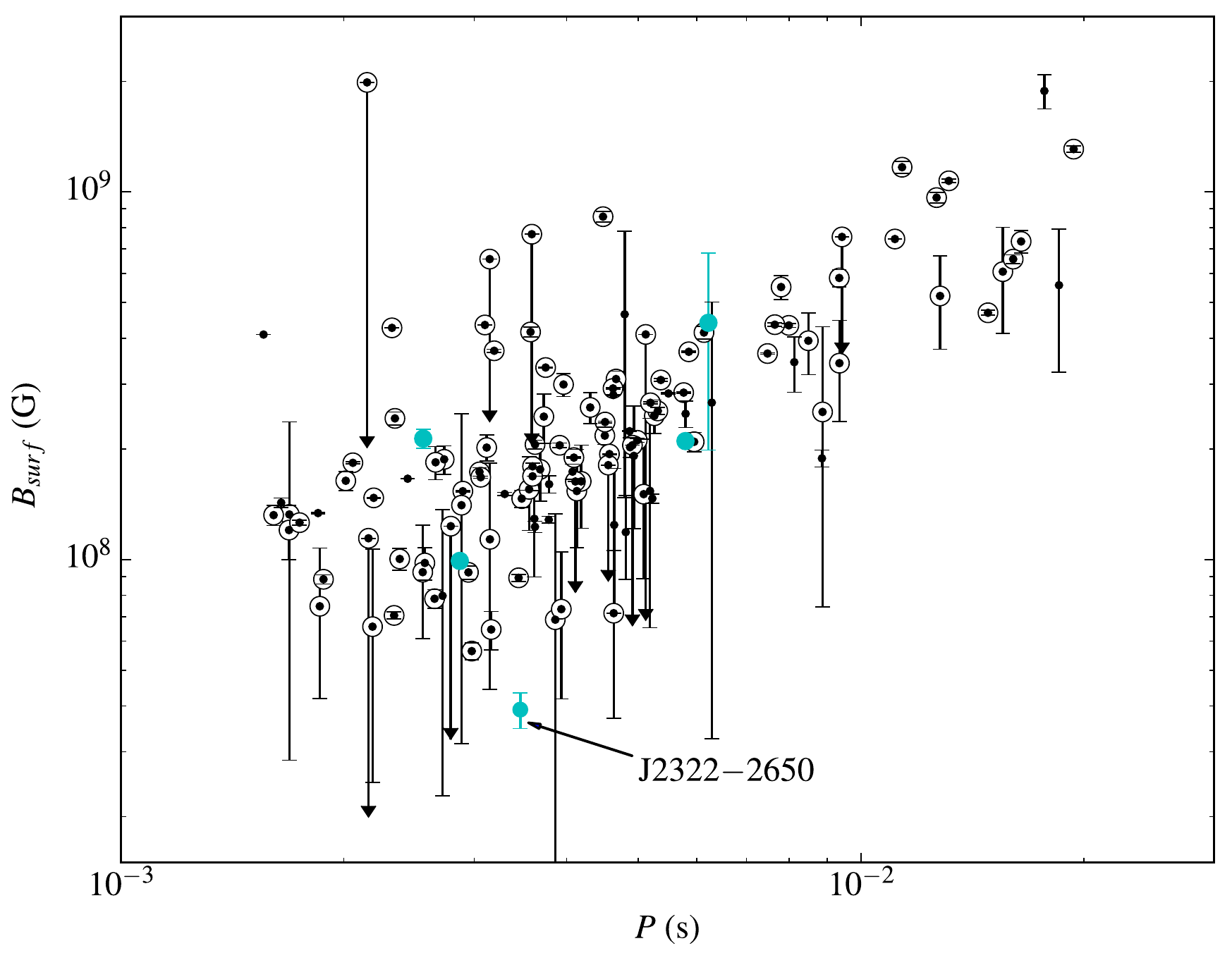}
 \caption{Intrinsic magnetic field strength versus period for field MSPs (data from the ATNF Pulsar Catalogue v1.56). 
 Black circles around dots indicate binary systems (with 1-$\sigma$ uncertainties), and dots alone indicate isolated systems.  1-$\sigma$ upper limits are used for pulsars with poorly constrained distance and proper motion measurements. The MSPs with planetary-mass companions are shown by the filled cyan circles. \psr\ is annotated, lying lower than MSPs with comparable periods, with 2-$\sigma$ uncertainties.  } 
 \label{fig:pvb}
\end{figure}  

Using the optimal value for $\dot P_{\rm int}$, we find $B_{\rm surf,i} = 4.0(5)\times10^7$\,G. 
Figure~\ref{fig:pvb} compares periods and magnetic field strengths of known MSPs, corrected for secular acceleration.  MSPs in globular clusters have been excluded due to the dominant effect of gravitational acceleration from their environments.  Note some field MSPs 
have negative period derivatives when corrected for the Shklovskii effect, largely due to contributions of the Galactic potential, and are therefore excluded from this Figure. \psr\ has the lowest intrinsic magnetic field strength of the remaining field MSPs. The other pulsars with planetary-mass companions have magnetic field strengths comparable with the other field MSPs with similar periods. 
As noted in Table~\ref{tab:par}, the intrinsic spin-down luminosity of \psr\ is $\dot E_{\rm int}  = 4.2(4)\times10^{32}$\,erg\,s$^{-1}$.

\subsection{Energetics} \label{ssec:lum} 
\begin{figure}
 \centering
 \includegraphics[width=\columnwidth]{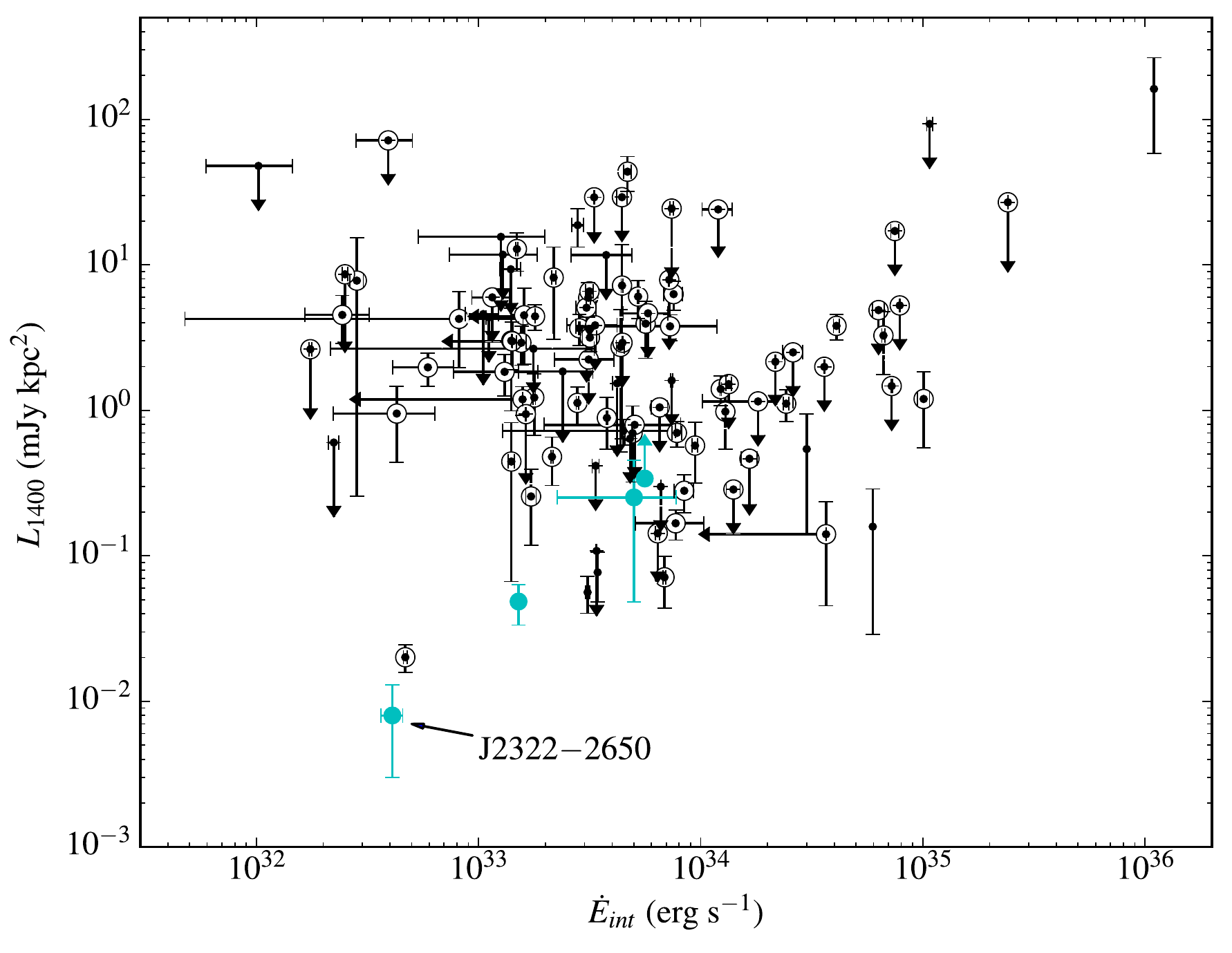}
 \caption{Radio luminosity versus intrinsic spin-down luminosity for MSPs in the field (data from the ATNF Pulsar Catalogue v1.56).  
 Black circles around dots indicate binary systems (with 1-$\sigma$ uncertainties), and dots alone indicate isolated systems.  1-$\sigma$ upper limits are used for pulsars with poorly constrained distance and proper motion measurements. The MSPs with planetary-mass companions are shown by the filled cyan circles. \psr\ is annotated, with the lowest radio luminosity. }
 \label{fig:lum}
\end{figure}

As noted in \S~\ref{ssec:time}, the mean flux density of \psr\ at 1400\,MHz is $S_{1400} = 0.16(2)$\,mJy, so the radio luminosity of the source is $L_{1400} = 0.008(5)$\,mJy\,kpc$^2$ (using the parallax-derived distance). 
This, too, is highly dependent on the distance measure.  At the parallax-derived distance of 230\,pc, the luminosity is lower than all consistent published values\footnote{\label{note:lum}Again, using ATNF catalogue sources with consistent distance measurements, and measured flux density at 1400\,MHz; 1684 pulsars (120 MSPs)}. 
Figure~\ref{fig:lum} shows a comparison of radio luminosity and intrinsic spin-down luminosity for field MSPs with directly measured 1400\,MHz flux density and reliable distance measurements. 
We distinguish binary and isolated systems in the Figure, but note no obvious difference between these populations, or correlation between the quantities, in this comparison. Of the pulsars with planetary-mass companions, PSR\,J1311$-$3430 is not plotted in this Figure as the 1400\,MHz flux density has not been measured, and PSR\,\psre\ is plotted at the lower limit of the luminosity from the timing parallax (see \S~\ref{ssec:comp}). 

\subsection{Binary parameters}\label{ssec:bin} 
\begin{figure}
 \centering
 \includegraphics[width=\columnwidth]{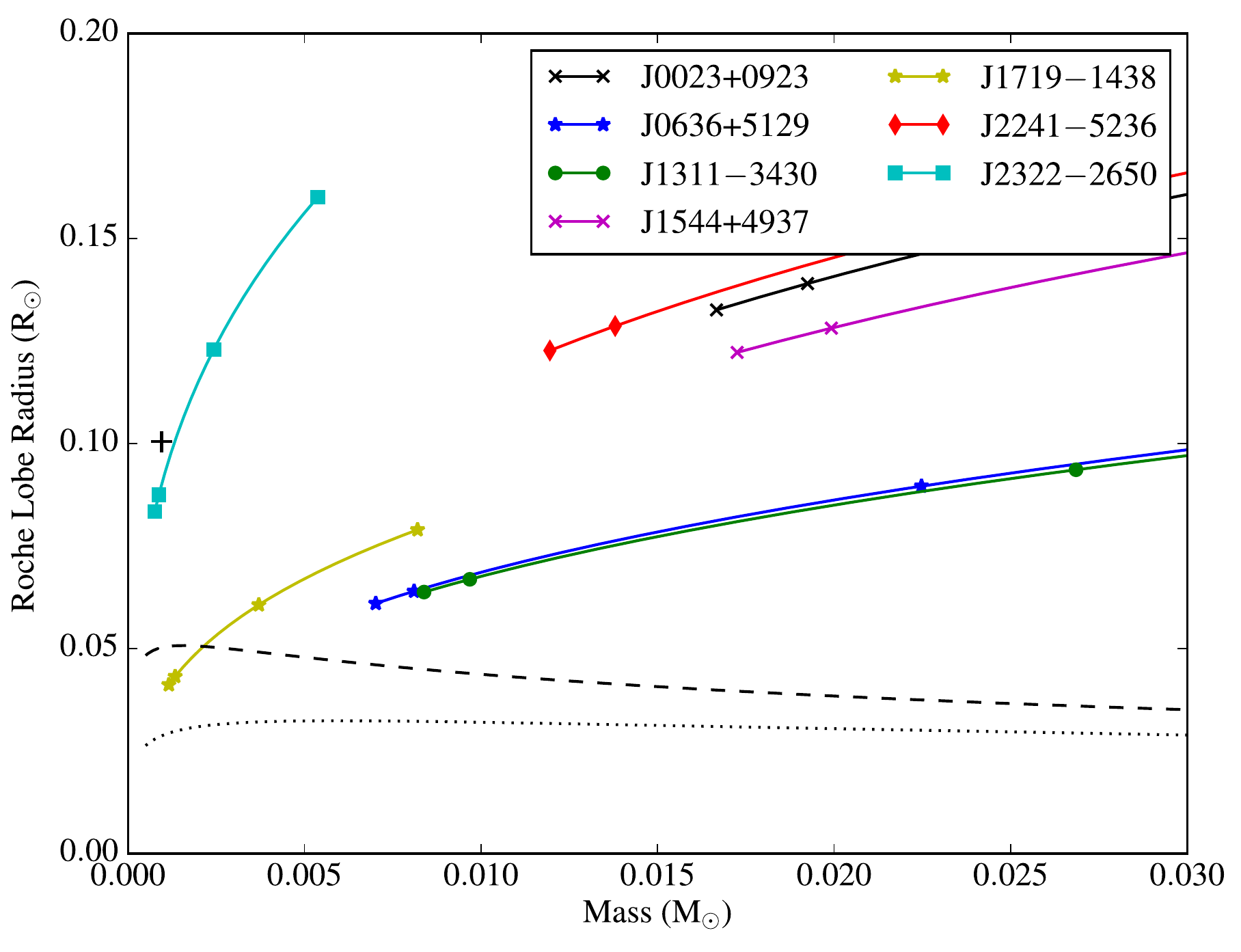}
 \caption{Plot of maximum Roche lobe radii versus companion mass for field MSPs with companions with very low masses ($x<0.04$\,lt-s and $m_{\rm min}<0.02$\,M$_{\odot}$).  
 The points from lowest to highest mass represent the maximum ($i = 90\degree$), median ($i = 60\degree$), 5\,percent ($i = 18.2\degree$), and 1\,percent ($i = 8.1\degree$) probabilities for the system inclination angle.  The dashed and dotted lines indicate the mass-radius relations for low-mass He and C white dwarfs, respectively, by Eggleton \citep{rmjn87}. 
 For reference, the mass and radius of Jupiter is marked with a cross.}  
 \label{fig:rlm}
\end{figure}

From the binary period, $P_{\rm b}= 0.322963997(6)$\,d, and projected semi-major axis, $x= 0.0027849(6)$\,lt-s, we find the mass function of \psr\ is $2.23\times10^{-10}$\,M$_{\odot}$, so the minimum companion mass is $M_{\rm c,min} = 0.000759$\,M$_{\odot}$, assuming a pulsar mass of $m_{\rm p} = 1.4$\,M$_{\odot}$. 
For lower inclination angles and higher pulsar masses, the companion mass increases, but remains below 0.01\,M$_{\odot}$ for $m_{\rm p}\leq2.0$\,M$_{\odot}$ and $i\geq8.1\deg$ (99\,percent probability given random system orientations).  
From the binary period, $P_{\rm b}$, we calculate the minimum density of the companion \citep{fkr85}: 
\begin{equation}
\rho = \frac{3\pi}{0.462^3GP_{\rm b}^2} = 1.83\,{\rm g\,cm}^{-3}. 
\end{equation}
In Figure~\ref{fig:rlm}, we 
plot the relation between the Roche lobe radius and the mass of the
companion for binary MSPs in the Galactic field with light companions
(i.e., having a minimum companion mass smaller than 0.02\,M$_{\odot}$). Each line 
in the plot covers the 99\,percent most probable orbital inclinations for any
given MSP binary. We note that the range of masses and radii for \psr\ is comparable to the mass and radius of Jupiter.

No post-Keplerian or higher-order binary parameters have been required in the parameter fits (see \S~\ref{sec:pars}). 
The advance of periastron, $\dot \omega$, cannot be included in the parameter fits, but the value from relativistic effects can be calculated assuming a pulsar mass of 1.4\,M$_{\odot}$, giving $\dot\omega_{\rm min}\approx1.6$\,deg\,yr$^{-1}$, which is not likely measurable due to the extremely low eccentricity of the orbit.  

We see no evidence for delays in the timing at superior conjunction of the pulsar.  This implies that there is no excess material in the system, or that the inclination of the system with respect to the line of sight prevents such material from affecting the delays of the pulsar signal.

\subsection{Multi-wavelength observations}\label{ssec:mwl}
We searched archives of {\it Fermi}, {\it Chandra}, and {\it XMM-Newton} missions for counterparts at other wavelengths.  No observations within 10\arcmin\ of the radio position were found in {\it Chandra} or {\it XMM-Newton} archives. 
The {\it Fermi} LAT 4-year Point Source Catalogue \citep{aaa+15} listed no sources within $30\arcmin$.  
An attempt to detect the pulsations using
our ephemeris and the entire Fermi dataset was not successful
(M.~Kerr private communication). 
From Figure 17 in \citet{aaa+13}, we estimate the upper limit on the flux density from 0.1 to 100\,GeV at a Galactic latitude of $b=-70\deg$ to be $\lesssim 3\times10^{-12}\,{\rm erg\,s^{-1}\,cm^{-2}}$, which would correspond to a luminosity of $L_\gamma\lesssim2\times10^{31}\,{\rm erg\,s^{-1}}$. The implied $\gamma$-ray efficiency\footnote{$\eta_\gamma$ is defined as the ratio of $L_\gamma$ and $\dot E_{\rm int}$.} is therefore $\eta_\gamma\lesssim 5\times10^{-2}$, which is consistent with MSPs detected in that energy range, as shown in \citet{aaa+13}. 

\psr\ is also undetected in a $\sim 1700$-s observation (PI J.~L.~Linsky, ROR 200461) performed on 20 Nov 1991 (UT 22:11) with the ROSAT PSPC \citep{pbf03} targeting HR8883, a star located $\approx19\arcmin$ from the radio position of \psr. We reanalyzed this ROSAT pointing using standard tools. In order to establish an upper limit to the observed X-ray flux, the analysis accounted for {\it (i)} the offset from the center of the field of view, and {\it (ii)} the expected low X-ray absorption column density toward the source (estimated using the Leiden/Argentine/Bonn Survey of Galactic HI; \citealt{kbh+05}). We also {\it (iii)} assumed a power-law spectrum, exploring photon indices around $-2$, which is often applied for inferring upper limits to the non-thermal X-ray emission from radio pulsars \citep[e.g.,][]{becker09}. The result was a $3\sigma$ upper limit to the unabsorbed X-ray flux of $\sim 2\times 10^{-13}\,{\rm erg~cm^{-2}s^{-1}}$ in the 0.1-2.4\,keV band. Since there is evidence for predominantly thermal X-ray emission from MSPs with intrinsic spin-down power ${\dot E}_{int}\lesssim 10^{35}\,{\rm erg~s^{-1}}$ \citep[see, e.g.,][]{kdpg12}, the consequences of the assumption of a Black Body spectrum were also explored. For surface temperatures in the range $0.5-5\times 10^6$ K (reflecting what is typically observed in the MSP sample), an upper limit on the unabsorbed X-ray flux similar to the one above was obtained. This limit corresponds to an isotropic X-ray luminosity $L_{\rm [0.1-2.4\,keV]}\lesssim 10^{30}\,{\rm erg~s^{-1}}(\frac{d}{\rm 230~pc})^2$ in the ROSAT PSPC band, where $d$ is the distance of the \psr\ binary, and the luminosity is scaled to the timing parallax distance. The implied upper limit to the X-ray efficiency\footnote{$\eta_X$ is defined as the ratio of $L_{\rm [0.1-2.4\,keV]}$ and ${\dot E}_{\rm int}$.} of the pulsar, $\eta_X\sim 2\times10^{-3}(\frac{d}{\rm 230~pc})^2$, agrees with what is seen in the bulk of the MSP population \citep[e.g.,][]{pccm02,becker09,kdpg12}. 

If we assume emission from the pulsar is heating the companion, we can estimate the expected blackbody temperature and optical brightness of the system.  We assume a certain geometry for the system: that the orbit is edge-on (the most likely and optimistic orientation for detection) and that the companion is tidally locked and filling its Roche lobe. As shown in \S~\ref{ssec:bin}, the system has an orbital period of $P_{\rm b} \approx 0.322964$\,d and projected semi-major axis of $x = a_1 \sin{i} \approx 0.002785$\,lt-s, and, therefore, the minimum companion mass is $m_{\rm c} \approx 0.000759$\,M$_{\odot}$. 
From Kepler's Third Law, because $m_{\rm c} \ll m_{\rm p}$, to a high degree of accuracy the system separation is  
\begin{equation}
a = 4.208\,{\rm R}_{\odot}\left(\frac{P_{\rm b}}{{\rm d}}\right)^{2/3}\left(\frac{m_{\rm p}}{{\rm M}_{\odot}}\right)^{1/3} = 2.2\,{\rm R}_{\odot}. 
\end{equation}
From this, the Roche lobe radius of the companion \citep{paczynski71} is 
\begin{equation}
R_{\rm L} = 0.462\,a\left(\frac{m_{\rm c}}{m_{\rm c}+m_{\rm p}}\right)^{1/3} = 0.083\,{\rm R}_{\odot}. 
\end{equation}
If the spin-down power, $\dot E\approx4.2\times10^{32}$\,erg\,s$^{-1}$, is emitted isotropically, 
the minimum blackbody temperature of the companion is $T_{\rm eff} \approx 2300$\,K.  
This would result in an apparent visual magnitude of $V\approx28$\,mag at system quadrature, 
and $V\approx27$\,mag at inferior conjunction of the pulsar. 
At the position ($l = 23.64\deg, b = -70.23\deg$) and estimated distance of the system, we estimate the absorption to be $E(B-V)\approx 0.01\pm0.03$\,mag, following \citet{gsf+15}\footnote{\url{http://argonaut.skymaps.info/}}, which is negligible. 
Given the low $T_{\rm eff}$ of the companion, we use the relations found by \citet{reed98} and \citet{cfb08} for cool M dwarfs to estimate $V-R\approx2$, implying a magnitude of $R\approx26$\,mag at quadrature ($R\approx 25$\,mag at inferior conjunction).  

\begin{figure}
 \centering
 \centering
 \includegraphics[width=0.85\columnwidth]{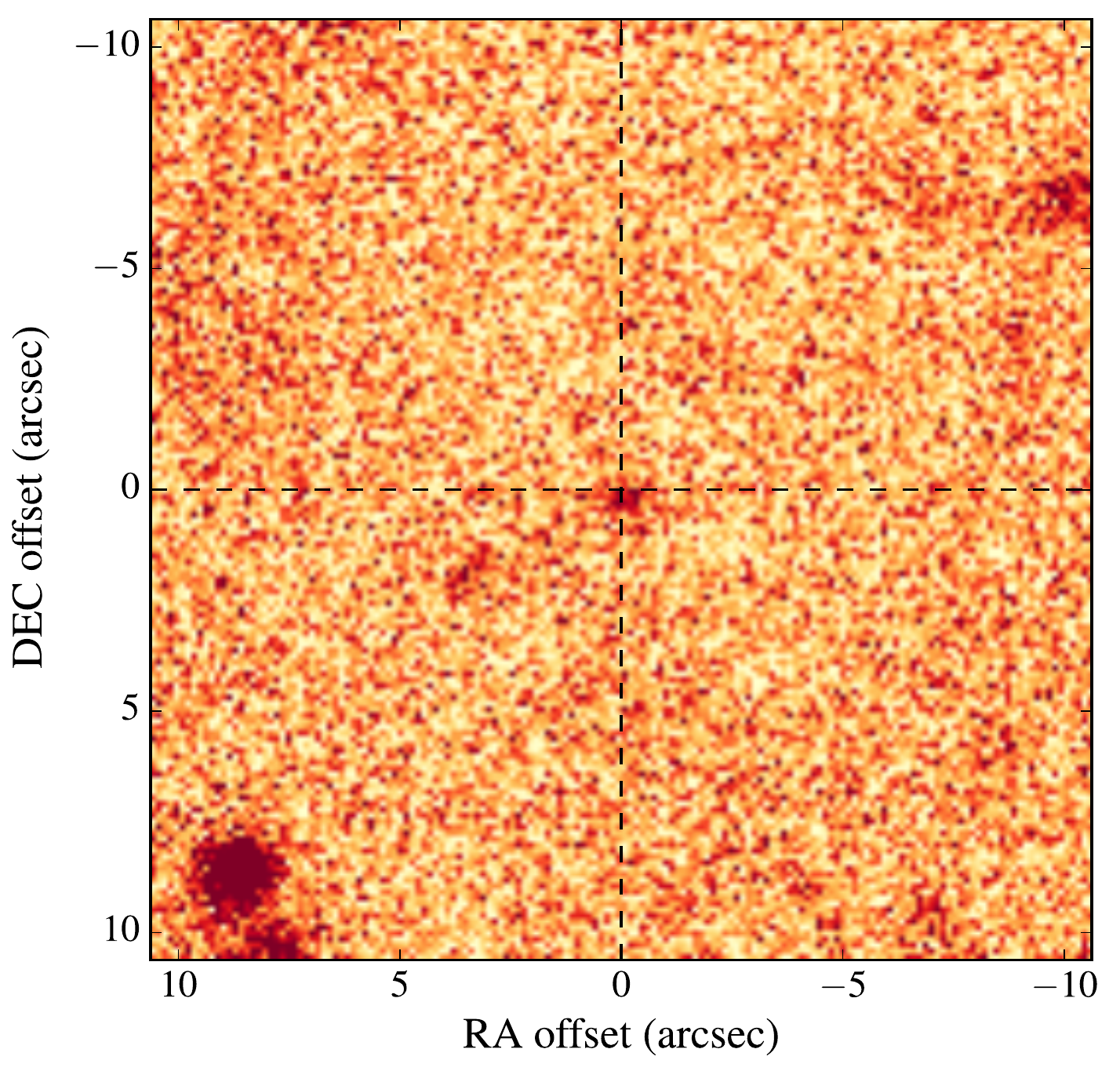}
 \caption{An image from the (summed) 1500\,s observation with the Keck DEIMOS instrument in $R$-band.  The axes indicate the offset from the radio pulsar position, with the black dashed lines denoting zero offset in RA and Dec. The centroid of the possible counterpart is $\sim0.6\arcsec$ from the radio position. }
 \label{fig:keck}
\end{figure}

In an attempt to detect the companion, we took three 500\,s exposures with the Keck DEep Imaging Multi-Object Spectrograph (DEIMOS) instrument in $R$-band on 2016 Sept.~8, as shown in Figure~\ref{fig:keck}.  
We detect an optical source $0.6\arcsec$ from the radio position (with $0.3\arcsec$ uncertainty in the astrometry) with an apparent $R$-band magnitude of $26.4\pm0.2$, where the uncertainty is given by Source Extractor \citep{ba96}, and we estimate systematic error of up to 0.2\,mag may also be present. 
The seeing of the observation was $\approx1.1\arcsec$, estimated from stellar sources in the field of view, and the limiting magnitude is $\approx25.8$\,mag (completeness limit). 
We estimate the probability of a random alignment of the radio position with an optical source as 4\,percent for sources down to $R\approx 25.8$\,mag. Therefore, the association of the optical and radio sources is approximately at the 2-$\sigma$ confidence level. 
The observation commenced at orbital phase $\approx0.75$ (inferior conjunction of the pulsar) for which our estimate of the blackbody emission results in an $R$-band magnitude of $\approx25$. 
Further observations at a range of orbital phases and better astrometry will ultimately determine whether the 26.4\,mag source is indeed the planetary-mass companion to \psr.

\section{Discussion} \label{sec:disc}
\subsection{Population statistics}\label{ssec:pop}

One of the most striking properties of \psr\ is its low luminosity of 0.008(5)\,mJy\,kpc$^{2}$.
Low luminosity MSPs appear in surveys relatively rarely unless their Galactic population
is very large, and in this section we explore what fraction of the total MSP population
might resemble \psr, cognizant of the fact that we are basing our discussion on just one object. 

In order to compare the Galactic population of \psr-like pulsars to the Galactic MSP population, a thorough analysis of the selection biases in our survey is necessary. 
To this end, we use the \textsc{psrevolve} software\footnote{Developed by F.~Donea and M.~Bailes, based on work by D.~Lorimer. \url{http://astronomy.swin.edu.au/~fdonea/psrevolve.html} } 
to simulate a specified number of pulsars scattered throughout the galaxy with assumed 
spatial distribution and distribution of pulsar parameters (period, magnetic field, luminosity, intrinsic pulse width) and determine if each pulsar would be detected in certain pulsar surveys. 
As detailed in \citet{lbb+13}, the simulation distributes pulsars at Galactic positions using a radial Gaussian distribution with radial scale length $R=4.5$\,kpc, centred on the Galactic centre, and a vertical Gaussian distribution with scale height $z=500$\,pc. A database of the coordinates of observations for the HTRU hi-lat survey is used to define the survey region. If the simulated pulsar is within the survey region, the pulse width (accounting for scattering and dispersion smearing) is calculated. If the pulse width is less than the pulse period, the final condition for detection (the flux density of the pulsar compared with the flux density limit of the survey) is checked. For each simulation run, 
we simulated $\sim5\times10^5$ pulsars with the period, magnetic field strength, luminosity, and pulse width of \psr\ and checked how many were ``detected" in the hi-lat survey.  The number of pulsars simulated normalised by the number of pulsars ``detected" provides a scaling factor: the total number of \psr-like MSPs in the galaxy beaming towards Earth.  
The simulation does not take into account the evolution or formation of MSPs and binary systems, so this analysis merely estimates the current population of low-luminosity MSPs in the galaxy.

\textsc{psrevolve} uses the NE2001 DM-distance model, so, for consistency, we used as input the luminosity of \psr\ at the NE2001 distance of 320\,pc ($L_{1400} = 0.016$\,mJy\,kpc$^2$). 
Out of 20 runs, we found a mean scaling factor of $9\times10^4$ with a standard deviation of $5\times10^4$. 
We also simulated a brighter pulsar ($L_{1400}=0.16$\,mJy\,kpc$^2$; all other parameters identical to \psr) and found a scaling factor of $3.2(9)\times10^3$. The ratio of \psr-like MSPs to those with an order of magnitude higher luminosity is, therefore, $28\pm18$, which is consistent with the slope of the luminosity distribution found by \citet{lbb+13}: $({\rm d}\log N/{\rm d}\log L) = -1.45\pm 0.14$. 

We have also used the {\sc PsrPopPy} software package \citep{blrs14} to confirm our results, using identical spatial distributions and pulsar parameters.  With this software, we find a scaling factor of $(3.7\pm0.8)\times10^4$ for \psr-like MSPs and $(1650\pm90)$ for the higher luminosity MSPs, and therefore a ratio of $23\pm5$ for the populations. The {\sc psrevolve} software includes a rough model 
of the effects of RFI on the surveys, thereby decreasing the detection likelihood and increasing the scatter in the scaling factors for the runs. Neither simulation tool accounts for refractive scintillation, which would affect the rate of detection of nearby pulsars such as \psr. 
These results also do not reflect 
the expected uncertainties from Poisson statistics. With this analysis, we do not claim a significant determination of the total population of low-luminosity MSPs. Rather, the detection of even a single low-luminosity MSP may imply the existence of a population of such MSPs that may dominate the Galactic MSP population.

\subsection{Comparison with other known MSP binaries}\label{ssec:comp}

If MSPs that are recycled by stars that leave behind planetary-mass companions have systematically low radio luminosities, we might expect to see that reflected in other members of the population. Below we discuss this population of MSPs, with properties summarised in Table~\ref{tab:psrs}. 

\begin{table}
 \centering
 \caption{Pulsars with Low-Mass Companions - comparison of companion masses, and radio luminosities and efficiencies.\label{tab:psrs}}
 \begin{threeparttable}
 \begin{tabular*}{\columnwidth}{l c c c c}
  \hline 
  Pulsar & $M_{\rm C}$ & $L_{1400}$ & $\dot E_{\rm int}$  & $\epsilon$ \\
  & (M$_{\rm J}$) & (mJy\,kpc$^2$) & ($\times10^{33}$\,erg\,s$^{-1}$) & ($\times10^{-7}$)\\ \hline
  \psre\ & 7.2 & $>0.34$ & 5.60(6) & 4.6 \\
  \psrb\tnote{a} & 0.014 & $\approx0.3$ & 5(3) & 6.0 \\
  J1311$-$3430 & 8.6 & $\approx0.22$ & 41(3)\tnote{b} & 0.3 \\
  \psrd\ & 1.2 & $\approx0.049$\tnote{c}  & 1.52(5) & 2.3 \\
  \psr\ & 0.76 & $0.008(5)$  & 0.42(4) & 1.3 \\ \hline
 \end{tabular*}
 \begin{tablenotes}
  \item[a] The mass of planet A is listed for \psrb\ \\
  \item[b] No proper motion measured for J1311$-$3430; $\dot P_{\rm int}$ is approximated from the measured $\dot P$ \\
  \item[c] Radio luminosity using the YMW16 distance
 \end{tablenotes}
 \end{threeparttable}
\end{table}

Besides \psr, the only MSPs with planetary-mass companions are PSRs~J0636+5128, B1257+12, J1311$-$3430, and J1719$-$1438. We note that PSR B1620$-$26 also has a planetary-mass companion, but this system is in a globular cluster, and is therefore not directly comparable to \psr. 

\psre\ is a low-mass Black Widow systems (with no radio eclipses) with a 7.2\,M$_{\rm J}$ companion in a 1.60\,hr orbit (\citealt{slr+14}; therein referred to as J0636+5129\footnote{See footnote~\ref{note:psre}}). 
\citet{slr+14} discuss the possibility that this system was formed via runaway mass transfer and ablation, noting that there is no sign of excess material in the orbit from radio observations. 
The MSP has a mean flux density of $S_{1400} = 0.69$\,mJy and a lower limit on the distance of $>700$\,pc (from the NANOGrav 11-year Data Release; Arzoumanian et al.~in prep.), implying a radio luminosity of $>0.34$\,mJy\,kpc$^2$. 
The intrinsic spin-down luminosity of the pulsar is also not unusually low, at $5.60(6)\times10^{33}$\,erg\,s$^{-1}$. 

\psrb\ has 3 companions with masses in the range of $6.3\times10^{-5}$ to 1.35$\times10^{-2}$\,M$_{\rm J}$ with orbital periods of 25 to 98\,d \citep{wf92,kw03}.  
\citet{wolszczan97} conclude that the planets likely formed in an accretion disc during or after the transfer of matter from the original (stellar) companion onto the pulsar.  
The MSP has a mean flux density at 1400\,MHz of $\approx0.5$\,mJy (from P140 Parkes observations) %citation
and a parallax distance of $0.71(4)$\,kpc \citep{ysy+13}, which implies a radio luminosity of $\approx0.3$\,mJy\,kpc$^2$.  Similar to \psre, \psrb\ has an intrinsic spin-down luminosity of $\dot E_{\rm int} = 5(3)\times10^{33}$\,erg\,s$^{-1}$.  

J1311$-$3430 is another low-mass Black Widow system, first detected in a {\it Fermi} blind search, and has a 8.6\,M$_{\rm J}$ companion in a 1.57\,hr orbit \citep{pgf+12a}.
This MSP is in an eclipsing system where the pulsar is ablating its companion with a high-energy wind \citep{pgf+12a}, and may therefore be similar to the progenitors of \psre\ and \psrd.  
J1311$-$3430 was initially detected as a $\gamma$-ray source, and has only been detected in radio frequencies once \citep{rrc+13}, implying a flux density at that time of $S_{1400}\approx0.11(6)$\,mJy.  The DM-distance from this detection is $1.4(1)$\,kpc, which thus implies a radio luminosity of $\approx0.22$\,mJy\,kpc$^2$.  It has a significantly higher $\dot E_{\rm int}$ (approximated from the observed $\dot P$ as no proper motion has been measured) than the other low-mass pulsar systems, $4.1(3)\times10^{34}$\,erg\,s$^{-1}$, and the observed ablation of its companion is assumed to be a consequence of that energy loss. 

\psrd\ has a 1.2\,M$_{\rm J}$ companion in a 2.2\,hr orbit \citep{bbb+11a}.  
Like \psre, \psrd\ is a possible case of ablation due to an energetic wind \citep{bbb+11a}, although no excess material is now observable. 
For \psrd, there is some ambiguity in the distance from the DM, with YMW16 giving a value of 0.34(3)\,kpc and NE2001 giving $1.2(3)$\,kpc, and, as of May 2017, there is no published parallax value.  Combined with a flux density of $S_{1400}=0.42$\,mJy \citep{nbb+14}, the YMW16 (NE2001) distance estimate implies a radio luminosity of $\approx0.049$\,mJy\,kpc$^2$ ($\approx0.61$\,mJy\,kpc$^2$). 

In comparison with the other MSPs with planetary-mass companions, \psr\ most closely resembles \psre\ and \psrd, with similar companion masses and spin-down luminosities. 
The spin-down luminosity of \psr\ is lower than the mean of the MSP population (Fig.~\ref{fig:lum}), although the other MSPs with planetary-mass companions have more typical luminosities. 

It is interesting to note that \citet{bbb+13}, as well as previous studies by \citet{kxl+98} and \citet{bjb+97}, found that isolated MSPs and binary MSPs have different intrinsic luminosity functions, where isolated MSPs have lower luminosities on average than MSPs with companions.  They suggest the difference may reflect differing evolutionary histories for the two populations. However, in recent years, additional isolated MSPs with average or high luminosities have been discovered, such as PSRs\,J1747$-$4036 \citep{kcj+12,ckr+15} and J1955+2527 \citep{dfc+12}, which do not support a significant difference between the luminosities of the two populations. 
At the YMW16 distance, \psrd\ has a radio luminosity comparable to that of \psr, which is significantly less than the median radio luminosity for binary MSPs\footnote{See footnote~\ref{note:lum}}. 
\psrb\ differs significantly from the other pulsars with planetary-mass companions with a higher radio luminosity and multiple Earth-mass companions, and we expect this is due to a different formation scenario from the other systems (discussed below). 

As shown in Table~\ref{tab:psrs}, if we assume a beaming fraction of 1, 
the radio efficiencies of these pulsars, $\epsilon = L_{\rm r}/\dot E_{\rm int}$, where $L_{\rm r}$ is the radio luminosity at 1400\,MHz in erg\,s$^{-1}$, 
are comparable. The remainder of the known MSPs have a mean (median) efficiency of $4\times10^{-5}$ ($3\times10^{-6}$), with no significant difference between the distributions for isolated and binary MSPs.

\subsection{Formation scenarios}\label{ssec:form}
The possible formation scenarios for this system are, as above: planet formation around the main sequence progenitor to the pulsar, planet formation in a supernova fallback disc, and the evaporation or ablation of the original companion to an extremely low mass. 
Following \citet{mh01}, we consider it highly unlikely for the planet to have formed around the main sequence star and remained bound after the supernova event. 

\citet{kjhs15} have searched for periodicity in timing data for 151 young pulsars to place limits on the existence of planets around pulsars.  They find that planet formation within $\approx 1.4$\,AU is a rare phenomenon, so it is unlikely that the companion to \psr\ formed before the pulsar was recycled. 

An alternative scenario is the formation of the planetary-mass companion in the accretion disc from the original companion, and subsequent loss of the original companion. 
\citet{acrs82} define the accretion time as 
\begin{equation}
T_a \sim 1.4\times10^8\,{\rm yr}\,\left(\frac{M}{{\rm M}_{\odot}}\right)^{-2/3} \left(\frac{P}{{\rm ms}}\right)^{-4/3}I \dot m^{-1}_{17},
\end{equation}
where $I$ is the moment of inertia in units of $10^{45}\,{\rm g\,cm}^2$ and $\dot m_{17}$ is the accretion rate in units of $10^{17}\,{\rm g\,s}^{-1}$, yielding an accretion time of $\sim2\times10^7$\,yr for a pulsar mass of 1.4\,M$_{\odot}$, period of 3.46\,ms, and $\dot m_{17}=1\times10^{17}$\,g\,s$^{-1}$. 
\citet{hsc09} discuss the formation of Earth-mass planets in discs and find that such bodies form in $\sim10^7$\,yr, but their simulations do not form Jupiter-mass objects. Although we cannot completely reject this formation scenario, it is not our preferred model.

Given the similarities between \psr, \psre, and \psrd, it is possible that the planetary-mass companion we now observe is the remnant of the original companion after runaway mass transfer. Following \citet{acrs82}, the formation of a pulsar with a period of 3.46\,ms would require $\sim0.05$\,M$_{\odot}$ transferred from an evolved companion, assuming a final pulsar mass of 1.4\,M$_{\odot}$.  The minimum density of the companion is 1.83\,g\,cm$^{-3}$, which does not preclude a scenario where the original companion transferred material to the pulsar and was ablated by the pulsar wind, reducing the companion to a mass of $\approx 0.0008$\,M$_{\odot}$.  
\citet{srp92} describe an ablation scenario by relating the mass loss of the companion, $\dot M_2$, to the energy loss of the pulsar $L_{\rm p}$, as 
\begin{equation}
\dot M_2 \propto L_{\rm p}\left(\frac{R_2}{a}\right)^2, \label{eq:mass}
\end{equation}
where $R_2$ is the radius of the companion and $a$ is the separation.  We can compare \psr\ with PSR\,B1957+20, a Black Widow ablating its companion at a rate of $\dot M_2 \sim 3\times10^{16}$\,g\,s$^{-1}$ \citep{as94}. 
\psr\ has a lower spin-down luminosity, 
so Equation~\ref{eq:mass} implies a mass loss rate of just $\sim 10^{13}$\,g\,s$^{-1}$. At this rate, the companion to \psr\ would lose just 0.1\,M$_{\rm J}$ in $\sim10^9$\,yr. The high-luminosity, isolated MSP, PSR\,B1937+21, with its high spin-down luminosity $\dot E_{\rm int} = 1.1\times10^{36}$\,erg\,s$^{-1}$, with the same orbital parameters as \psr, would ablate the entire companion in only $\sim10^6$\,yr. Therefore, we speculate that \psr\ has a planetary-mass companion remaining due to its low spin-down luminosity, and that a more energetic pulsar with an identical original companion would destroy its companion and become isolated.

\section{Conclusions}\label{sec:con}

In this paper, we have presented the discovery of an MSP unlike other known MSPs: a nearby MSP, characterised by a low surface magnetic field strength and a low radio luminosity, with a low-density, planetary-mass companion.  A single observation of the system with the Keck DEIMOS instrument in $R$-band revealed a source of $R\approx26.4(4)$\,mag that is associated with the pulsar companion at the 2-$\sigma$ confidence level. 

If MSPs with planetary-mass companions have luminosities similar to \psr, they may dominate the galactic MSP population. Future surveys with telescopes like the SKA and FAST may reveal them.

\section*{Acknowledgements}
We thank I.~Andreoni for his assistance with reduction of the optical data, and P.~Esposito for his advice regarding the X-ray data analysis. We also thank the anonymous referee for useful comments that significantly improved the manuscript.  
This research was funded partially by the Australian Government through the Australian Research Council, grants CE170100004 (OzGrav) and FL150100148. 
MK's research is supported by the ERC Synergy Grant ``BlackHoleCam: Imaging the Event Horizon of Black Holes'' (Grant 610058). 
The Parkes radio telescope is part of the Australia Telescope National Facility which is funded by the Australian Government for operation as a National Facility managed by CSIRO. 
Pulsar research at the Jodrell Bank Centre for Astrophysics and the observations using the Lovell telescope are supported by a consolidated grant from the STFC in the UK. 
Some of the data presented herein were obtained at the W.M. Keck Observatory, which is operated as a scientific partnership among the California Institute of Technology, the University of California and the National Aeronautics and Space Administration. The Observatory was made possible by the generous financial support of the W.M. Keck Foundation. 
This work used the gSTAR national facility which is funded by Swinburne and the Australian Government's Education Investment Fund. 
This research made use of Astropy, a community-developed core Python package for Astronomy (Astropy Collaboration, 2013), and the Matplotlib package \citep[v1.5.1;][]{hunter07}.

\bibliographystyle{mn2e}
\bibliography{psr}   %%%%%%%%%  File updated Nov. 2017

% Don't change these lines
\bsp	% typesetting comment
\label{lastpage}
\end{document}